\documentclass[12pt]{article}
\begin{document}
\renewcommand{\thefootnote}{\fnsymbol{footnote}}
\centerline{\large\bf 
Exact form factors for the Josephson tunneling current   
}
\centerline{\large \bf and relative particle
number fluctuations in a model    
}
\centerline{\large\bf
of  two coupled Bose-Einstein condensates}

~~\\~~\\
\centerline{J. Links and H.-Q. Zhou}
~~\\
\centerline{Centre for Mathematical Physics,} 
\centerline{School of Physical Sciences} 
\centerline{The University of Queensland,
                      4072,} 
\centerline{Australia}

\vspace{10pt}

\begin{abstract}
Form factors are derived for a model describing the coherent Josephson
tunneling between two coupled Bose-Einstein condensates. This is
achieved by studying the exact solution of the 
model in the framework of the algebraic Bethe ansatz. In this approach
the form factors are expressed through determinant representations which are
functions of the roots of the Bethe ansatz equations. 
\end{abstract}

\begin{flushleft} 
\end{flushleft} 

\vfil\eject 
                     


\def\a{\alpha}
\def\b{\beta}
\def\d{\dagger}
\def\e{\epsilon}
\def\g{\gamma}
\def\K{\kappa}
\def\l{\lambda}
\def\o{\omega}
\def\t{\theta}
\def\s{\sigma}
\def\D{\Delta}
\def\L{\Lambda}
\def\ap{\approx} 


\def\beq{\begin{equation}}
\def\eeq{\end{equation}}
\def\bea{\begin{eqnarray}}
\def\eea{\end{eqnarray}}
\def\ba{\begin{array}}
\def\ea{\end{array}}
\def\no{\nonumber}
\def\le{\langle}
\def\re{\rangle}
\def\lt{\left}
\def\rt{\right}
\def\o{\omega}
\def\d{\dagger}
\def\nn{\nonumber} 
\def\j{{ {\cal J}}}
\def\n{{\hat n}}
\def\N{{\hat N}}
\newcommand{\reff}[1]{eq.~(\ref{#1})}

\vskip.3in

The experimental realisation of Bose-Einstein condensates in atomic
alkali gases is generating a great deal of theoretical activity in order to
understand the nature  of coherent Josephson tunneling  
between coupled systems. A simple two-mode Hamiltonian which has been 
widely studied (see \cite{leggett} for a review)   
takes the form
\bea
H&=& \frac {1}{8} K \n^2 - \frac {\Delta \mu}{2} \n
 -\frac {{\cal {E}_J}}{2} (a_1^\dagger a_2 + a_2^\dagger a_1).
\label{ham} \eea
where $a_1^\dagger, a_2^\dagger$ denote the single-particle boson creation
operators, $\N_1 = a_1^\dagger a_1$, 
$\N_2 = a_2^\dagger a_2$ are number operators and $\n=\N_1-\N_2$ is
the  relative particle 
number operator. Note that the total number operator 
$\N=\N_1+\N_2$ provides
a good quantum number since $[H,\,{\hat N}]=0.$  

While this model has been studied using 
a variety of techniques such as 
Gross-Pitaevskii states \cite{leggett}, mean-field theory
\cite{milburn},  quantum phase model \cite{qpm} and an exact quantum phase
model \cite{drummond}, it also admits an exact solution
in the framework of the algebraic Bethe ansatz, given
in \cite{enollskii} in the guise of the ``discrete self-trapping dimer''
model, which
has been largely unexplored. In \cite{us}, we have used the exact solution
to determine the asymptotic behaviour of the energy spectrum 
in the Fock ($N<<K/{\cal {E}}_J$)
and Rabi ($N^{-1}>>K/{\cal {E}_J}$) regions. 
It was also shown that asymptotic 
thermodynamic properties can be deduced for
the Fock region at low temperature and the Rabi region for all
temperatures. Here, we will limit
ourselves to the case $\Delta \mu=0$ and continue our
analysis of  the Bethe ansatz 
solution to yield explicit exact form factors for the 
Josephson tunneling current 
\beq \j=i(a^{\d}_1a_2-a^{\d}_2a_1), \label{current} \eeq  
as well as the relative particle number $\n$ 
and $\n^2$, which are
applicable for all couplings. This opens an avenue to explore the
behaviour of the Hamiltonian in the cross over between the   
Fock, Rabi and Josephson ($N^{-1}<<K/{\cal {E}}_J<<
N$) regions. Recall that in
Josephson's original work on macroscopic superconductors \cite{j}, the
tunneling current is a manifestation of the relative phase of the wave
functions of two
superconductors separated by an insulating barrier. As phase and
particle number are canonically conjugate quantum variables, 
the quantum fluctuations of the relative particle number are of primary
importance in understanding the physics in the present model, 
given that there are technical difficulties which prevent a simple 
definition for the phase variable \cite{leggett,phase}. 
Our results for the form factors provide an initial step towards the
calculation of these fluctuations. 
The method that we will adopt follows that proposed for
the form factors of the Bose gas with delta-function 
interactions and the one-dimensional
Heisenberg model \cite{korepin,maillet}. 
This yields determinant representations for the form
factors which are functions of the roots of the Bethe ansatz equations 
that arise from the exact solution. Furthermore, it is straightforward
to express these results in a time-dependent form.

First we will review the basic features of the exact solution of  
(\ref{ham}) via the algebraic Bethe ansatz, as discussed in 
\cite{enollskii}. The theory of exactly solvable quantum systems in this 
setting relies on the existence of a solution of the 
Yang-Baxter equation
\beq
R _{12} (u-v)  R _{13} (u)  R _{23} (v) =
R _{23} (v)  R _{13}(u)  R _{12} (u-v). 
\label{ybe} \eeq
Here $R_{jk}(u)$ denotes the matrix on $V \otimes V\otimes V$ acting on the
$j$-th and $k$-th spaces and as the identity on the remaining space.
The $R$-matrix solution may be viewed as the structural constants for the
Yang-Baxter algebra generated by the monodromy matrix $T(u)$
\beq
R_{12}(u-v) T_1(u) T_2(v)=
T_2(v) T_1(u)R_{12}(u-v). \label{YBA}
\eeq
For a given $R$-matrix, there are a variety of realisations of 
the Yang-Baxter algebra. For the $su(2)$ invariant $R$-matrix,
\beq
R(u) = \left ( \begin {array} {cccc}
1&0&0&0\\
0&b(u)&c(u)&0\\
0&c(u)&b(u)&0\\
0&0&0&1\\
\end {array} \right ),
\label{rm} \eeq
with the rational functions $b(u)=u/(u+\eta)$ and 
$c(u)=\eta/(u+\eta)$, there is a realisation of the
Yang-Baxter algebra in terms of canonical boson operators which reads
\cite{kt} 
$$
L(u) = \left ( \begin {array} {cc}
u+\eta \N &a\\
a^\dagger & \eta ^{-1}\\
\end {array} \right ), 
$$
such that 
$$R_{12}(u-v)L_1(u)L_2(v)=L_2(v)L_1(u)R_{12}(u-v). $$ 
The co-multiplication behind the Yang-Baxter
algebra allows us to choose the monodromy matrix
\bea
T(u)& =&  L_1(u+\omega) L_2(u-\omega) \no \\
&=& \pmatrix{ (u+\o+\eta \N_1)(u-\o+\eta \N_2)+a_2^{\d}a_1 &
(u+\o+\eta \N_1)a_2+\eta^{-1}a_1 \cr 
(u-\o+\eta \N_2)a_1^{\d}+\eta^{-1}a^{\d}_2 & a_1^{\d}a_2+\eta^{-2} }
\no \\ 
&&\label{real} \eea 
Defining the transfer matrix through 
$t(u) = {\rm tr}\, (T(u))$, it follows from (\ref{ybe}) that the 
transfer matrices commute for different values of the spectral parameter;
viz. 
$$[t(u),\,t(v)]=0~~~\forall u,v.$$ 
In the present case, we have explicitly 
\bea 
t(u) &=& u^2 + u \eta \N 
+ \eta^2 \N_1\N_2 + \eta \omega\n
+a_2^\dagger a_1+ a_1^\dagger a_2 + \eta^{-2} - \omega^2.
\nn \eea 
Then 
$$ t^{\prime}(0)= \left.\frac{dt}{du}\right|_{u=0}= 
\eta\N $$
and it is easy to verify that the Hamiltonian is related with the transfer
matrix $t(u)$ by 
$$
H=-\K \left (t(u) -\frac {1}{4} (t'(0))^2-
u t'(0)-\eta^{-2}
+\omega^2 -u^2\right),  
$$
where the following identification has been made for the coupling constants 
\bea
\frac {K}{4} &=&  \frac {\K \eta^2}{2}, \no\\
\frac {\Delta \mu}{2} &=&  -\K \eta \omega, \no\\
\frac {\cal {E}_J}{2} &=&  \K . \no
\eea

The solution of (\ref{ham}) via the algebraic Bethe ansatz is obtained by 
utilizing the commutation relations of the Yang-Baxter algebra. 
Setting
\beq
T(u) = \left ( \begin {array} {cc}
A(u)&B(u)\\
C(u)&D(u)
\end {array} \right ),\label{mono}
\eeq
we have from the defining relations (\ref{YBA}) that  
\bea
& & [A(u), A(v)] =
    [D(u), D(v)] = 0, \no\\
& & [B(u), B(v)] =
    [C(u), C(v)] = 0, \no\\
& & A(u)C(v) =
\frac {u-v+\eta}{u-v} C(v) A(u)-\frac {\eta}{u-v}
C(u) A(v), \no\\
& & D (u) C(v) =
\frac {u-v-\eta}{u-v} C(v) D(u) +\frac {\eta}{u-v}
C(u) D(v). \label{rels}   
\eea
An explicit representation of (\ref{mono}) is obtained from (\ref{real})
with the identification
\bea 
A(u)&=& (u+\o+\eta N_1)(u-\o+\eta N_2)+a_2^{\d}a_1 \nn \\
B(u)&=&
(u+\o+\eta N_1)a_2+\eta^{-1}a_1 \nn \\
C(u)&=&
(u-\o+\eta N_2)a_1^{\d}+\eta^{-1}a^{\d}_2 \nn \\
D(u)&=& a_1^{\d}a_2+\eta^{-2}. \nn \eea  
Choosing the state $\left|0\right> = \left|0 \right> _1  
\left|0 \right>_2 $
as the pseudovacuum, the eigenvalues $a(u)$ and $d(u)$
of $A(u)$ and $D(u)$ on $\left|0\right>$ are 
\bea
a(u)  &=& (u+\omega)(u-\omega),\no\\
d(u)  &=& \eta^{-2}.\nn
\eea
Next choose the Bethe
state 
\beq \left|\vec v\right>\equiv\left|v_1,...,v_N\right>
= \prod ^N_{i =1} C(v_i) \left|0 \right>. \label{state} \eeq 
Note that because $[C(u),\,C(v)]=0$, the ordering is not important in
(\ref{state}). It is also clear that these states are eigenstates of $\hat N$
with eigenvalue $N$. The approach of the algebraic Bethe ansatz is to
use the relations (\ref{rels}) to determine the action of $t(u)$ on
$\left|\vec v\right>$. The result is 
\bea  
t(u) \left|\vec v\right> 
&=& \L (u,\,\vec v) 
\left|\vec v\right>  \nn \\
&&~~-\left(\sum_i^N\frac{\eta a(v_i)}{u-v_i}\prod_{j\neq i}^{N}
\frac{v_i-v_j+\eta}{v_i-v_j}
\right)\left|v_1,...v_{i-1},u,v_{i+1},...,v_N\right> \nn \\ 
&&~~+\left(\sum_\a^N\frac{\eta d(v_i)}{u-v_i}\prod_{j\neq i}^{N}
\frac{v_i-v_j-\eta}{v_i-v_j}
\right)\left|v_1,...v_{i-1},u,v_{i+1},...,v_N\right> \label{osba}  
\eea 
where
\beq 
\L(u,\,\vec v) = a(u) \prod ^N_{i=1} 
\frac {u-v_i+\eta}
{u-v_i}+
d(u) \prod ^N_{i=1} \frac {u-v_i-\eta}
{u-v_i}.  
\label{tme} \eeq 
The above shows that $\left|\vec v\right>$ becomes an eigenstate of the
transfer matrix with eigenvalue (\ref{tme}) whenever
the Bethe ansatz equations (BAE)
\beq
\eta^2 (v^2_i -\omega^2)=
\prod ^N_{j \neq i}\frac {v_i -v_j - \eta}{v_i -v_j +\eta}
\label{bae} \eeq
are satisfied. As $N$ is the total number of bosons,
we expect $N+1$ solutions of the BAE. 
Note that in the derivation of the BAE it is required that $v_i\neq v_j\,
\forall\,i,\,j.$ 
For example, the solution
$$v_j=\pm(i)^{(N+1)}\sqrt{\eta^{-2}+\o^2},~~~~\forall j$$
of (\ref{bae}) is invalid, except when $N=1$.  
This is a result of the Pauli Principle for Bethe 
ansatz solvable models as developed in \cite{ik82} for the
Bose gas. 
We will not reproduce the proof for the present case, as it follows
essentially the same lines as \cite{ik82}.  
For a given valid solution of 
the BAE, the energy of the
Hamiltonian is obtained from the transfer matrix eigenvalues 
(\ref{tme}) and reads 
\beq E(\vec v)
=-\K\left(\eta^{-2}\prod_{j}^N\eta^2(v_{j}-\omega+\eta)(v_{j}+\omega)
-\frac{\eta^2N^2}{4}
-\eta\o N-\eta^{-2}\right). \label{nrg}   \eeq  

A consequence of the above construction is that form factors, and consequently 
correlation functions, can be computed for this model. The method we
employ follows that used  for the Bose gas and 
one-dimenisonal 
Heisenberg chain \cite{korepin,maillet}. 
The reason we can adopt this approach is because 
the solution of these  models and (\ref{ham}) are based
on the same $R$-matrix (\ref{rm}). Below we will 
restrict our attention to $\o=0$ (for reasons which will become clear) 
and give explicit results 
for $\n,\,\n^2$ and $\j$.

A curiosity of this model is 
that the representation of the Yang-Baxter algebra is non-unitary; viz.
$$C^{\d}(u)\neq B(u). $$ 
In the case when $\o=0$ it is, however, equivalent to a unitary representation
since in this instance we have 
$$C^{\d}(u)=P.B(u).P $$ 
where $P$ is the permutation operator.  The permutation operator
is defined by the action on the Fock basis
$$P.(a^{\d}_1)^j(a^{\d}_2)^k\left|0\right>
=(a^{\d}_1)^k(a^{\d}_2)^j\left|0\right>. $$ 
Note that for $\o=0$ (which will be assumed hereafter), we have  
$$[P,\,H]=0$$ 
which means that the energy eigenstates 
are also eigenstates of $P$. 
Moreover,
$P^2=I$ shows that $P$ has eigenvalues $\pm 1$. 

There exists a  formula due to Slavnov \cite{slavnov} 
for the scalar product of states 
obtained via the algebraic Bethe ansatz for the $R$-matrix (\ref{rm}),
which when applied to this model is 
\bea S(\vec v:\,\vec u)&=&\left<0\right|B(v_1)...B(v_N)C(u_1)...C(u_N)
\left|0\right> \nn \\
&=&\frac{{\rm det} F(\vec u:\,\vec v)}{\prod_{j>k}(u_k-u_j)
\prod_{\a<\b}(v_\b-v_\a)}
\nn \eea 
where 
$$F_{ab}=\frac{\eta^{-1}}{u_b-v_a} \left(u^2_b \prod^N_{m\neq a}(v_m-u_b-\eta)
-\eta^{-2}\prod^N_{m\neq a}(v_m-u_b+\eta)\right), $$
the parameters $\{v_\a\}$ satisfy the Bethe ansatz equations (\ref{bae})
and $\{u_j\}$ are arbitrary. 
Note that when $\left|\vec u\right>=\left|\vec v\right>$ we need to take
a limit for the diagonal entries which gives 
\bea F_{aa}
&=&-2\eta^{-2}v_a\prod^N_{l}(v_l-v_a-\eta)
-\eta^{-4} \prod^N_{l}(v_l-v_a+\eta)\sum^N_{m\neq a} \frac{1}{v_m-v_a+\eta}
\nn \\
&&~+\prod^N_{l}(v_l-v_a-\eta)\sum^N_{m\neq
a}\frac{\eta^{-2}v_a^2}{v_m-v_a-\eta}.
\nn\eea 
Consider 
\bea S(\vec v:\,\vec v)&=&\left<0|B(v_1)...B(v_N)C(v_1)...C(v_N)|0\right> 
\nn \\  
&=&\left<0|PC^{\d}(v_1)...C^{\d}(v_N)P
C(v_1)...C(v_N)|0\right> \nn \\
&=&\e(\vec v)\left<\vec v|\vec v\right> \nn \eea 
where $\e(\vec v)=\pm 1$ is the eigenvalue of $P$; viz. 
$$P\left|\vec v\right>=\e(\vec v)\left|\vec v\right>. $$ 
This quantity can be determined to be given by 
$$\e(\vec v)=\prod_{j}^N\eta v_j. $$ 
Hence, from the Slavnov formula, the norms of the eigenstates 
\bea ||\vec v||&=&\left<\vec v|\vec v\right>^{1/2} \nn \\
&=&\left|S(\vec v:\vec v)\right|^{1/2} \nn \eea   
are obtained directly.

Let us define
$$\chi=A(0)-D(0)=\eta^2N_1N_2 +i\j -\eta^{-2} $$
where $\j$ is defined by (\ref{current}). 
In analogy with (\ref{osba}) we find
\bea
&&\chi \left|\vec u\right>\nn \\
&&~~= \theta (\vec u)
\left|\vec u\right>  \nn \\
&&~~+\sum_i^N\frac{\eta a(u_i)}{u_i}\prod_{j\neq i}^{N}
\frac{u_i-u_j+\eta}{u_i-u_j}
\left|u_1,...u_{i-1},0,u_{i+1},...,u_N\right> \nn \\
&&~~+\sum_i^N\frac{\eta d(u_i)}{u_i}\prod_{j\neq i}^{N}
\frac{u_i-u_j-\eta}{u_i-u_j}
\left|u_1,...u_{i-1},0,u_{i+1},...,u_N\right> \nn 
\eea
where
\bea
\theta(\vec u) &=& a(0) \prod ^N_{i=1}
\frac {u_i-\eta}
{u_i}-
d(0) \prod ^N_{i=1} \frac {u_i+\eta}
{u_i} \nn \\
&=&-\eta^{-2}\prod_{i=1}^N \frac{u_i+\eta}{u_i}. \nn
\eea
Using the Slavnov formula we can calculate the form factors for $\chi$. Suppose
that $\left|\vec v\right>$ is an eigenstate of the Hamiltonian. Then 
\bea
&&\left<\vec v|\chi |\vec u\right>\nn \\
&&~~= \theta (\vec u)
\e(\vec v)S(\vec v:\,\vec u)  \nn \\
&&~~+\sum_i^N\frac{\eta a(u_i)}{u_i}\prod_{j\neq i}^{N}
\frac{u_i-u_j+\eta}{u_i-u_j}
\e(\vec v)S(\vec v:u_1,...u_{i-1},0,u_{i+1},...,u_N) \nn \\
&&~~+\sum_\a^N\frac{\eta d(u_i)}{u_i}\prod_{j\neq i}^{N}
\frac{u_i-u_j-\eta}{u_i-u_j}
\e(\vec v)S(\vec v:\,u_1,...u_{i-1},0,u_{i+1},...,u_N). \nn
\eea
 
Now we let $\left|\vec u\right>$ be an eigenstate of the Hamiltonian
(\ref{ham}).
The above formula can be simplified considerably (cf. \cite{maillet}) 
\bea
&&\left<\vec v|\chi |\vec u\right>\nn \\
&&~=\frac{-\eta^{N-2}\prod_{\a}(v_\a+\eta)}
{\prod_{j>k}(u_k-u_j)\prod_{\a<\b}(v_\b-v_\a)}
{\rm det} \left(F(\vec v:\,\vec u)-2\eta^{-3}\e(\vec u)\e(\vec v)Q(\vec v:\, 
\vec u)\right) \nn \\
&&  \label{ff}  \eea
where $Q(\vec v:\, \vec u)$ is a rank one matrix with entries
$$Q_{ab}=\frac{\prod_{j}(u_j-u_b+\eta)}{v_a(v_a+\eta)}. $$ 
This is our main result. We remark that because the basis states are
also Hamiltonian eigenstates, it is straightforward to write down the
time-dependent form factors 
\beq \left<\vec v|\chi(t)|\vec u\right>=\exp(-it\left(E(\vec v)-E(\vec u)
\right))\left<\vec v|\chi|\vec u\right> \label{tdff} \eeq  
where the energies $E(\vec v)$ are given by (\ref{nrg}). 
Remarkably, from eqs. (\ref{ff}, \ref{tdff}) all the time-dependent
form factors for
$\n,\,\n^2$ and $\j$ can be obtained. This is achieved by exploiting the
symmetry of the Hamiltonian under $P$, which we will now explain.

The following corollary is easily proved. If $\e(\vec v)\neq \e(\vec u)$
then  
$$\left<\vec v|\N_1\N_2|\vec u\right>=0. $$ 
If $\e(\vec v)=\e(\vec u)$ then 
$$\left<\vec v|\j|\vec u\right>=0. $$ 
The result follows from the observation
\bea P\N_1\N_2&=&\N_1\N_2P, \nn \\
P\j&=&-\j P. \nn \eea 
 We now find that 
$$\left<\vec v|\N_1\N_2|\vec u\right> = 
\eta^{-2}\left<\vec v|\chi(t)|\vec u\right>
+\eta^{-4}\left<\vec v|\vec u\right> $$ 
if $\e(\vec v)=\e(\vec u)$, and is zero otherwise. Also  
$$\left<\vec v|\j|\vec u\right> = -i\left<\vec v|\chi(t)|\vec u\right>
$$
if $\e(\vec v)\neq \e(\vec u)$, and is zero otherwise. 

The above shows that the form factors for $\j$ are obtained directly from those 
of $\chi(t)$. Those for $\n^2$ also follow, since we have 
$\n^2={\hat N}^2-4\N_1\N_2$ and the states (\ref{state}) are automatically 
eigenstates
of $\hat N$ with eigenvalue $N$. Thus 
$$\left<\vec v|\n^2|\vec u\right>=N^2\left<\vec v|\vec u\right> 
-4\left<\vec v|\N_1\N_2|\vec u\right>. $$ 
To obtain the form factors for $\n$, we use the fact that 
$\j$ is the time derivative of $\n$, so  
$$\j=\frac{i}{\K  }[\n,\,H]$$ 
which gives 
$$\left<\vec v|\n|\vec u\right> =\frac{i\K}
{E(\vec v)-E(\vec u)} \left<\vec v|\j|\vec u\right>. $$ 

In principle, the correlation functions 
$$\left<\theta\right>_\Psi=\frac{\left<\Psi|\theta|\Psi\right>}
{\left<\Psi|\Psi\right>}$$ where $\theta=\n,\,\n^2$ or $\j$, and 
$\left|\Psi\right>$ is an  arbitrary state, can be expressed
in terms of the form factors through completeness relations. 
In particular, for a given $\left|\Psi\right>$ 
the quantum fluctuations of the relative number operator
$$\delta(\Psi;\n)=\left<\n^2\right>_\Psi
-\left<\n^{\phantom{2}}\right>_\Psi^2 $$
can be computed from these results. 

In conclusion, we have shown that the algebraic Bethe ansatz solution of 
(\ref{ham}) provides a means to calculate form factors, and in turn 
correlation functions, for the model. 
We have yielded explicit results for the 
case $\Delta\mu=0$. 
An outstanding task is to extend these results to the case 
$\Delta\mu\neq 0$, which presents a challenging problem because 
of the non-unitary representation of the Yang-Baxter algebra in 
this instance, and the breaking of symmetry under $P$. 

\vskip.3in
We thank  R.H. McKenzie for stimulating discussions and 
the Australian Research Council for financial support.


\begin{thebibliography}{99}
\bibitem{leggett}A.J. Leggett, Rev. Mod. Phys. {\bf 73} (2001) 307. 
\bibitem{milburn} G.J. Milburn, J. Corney, E.M. Wright and D.F. Walls,
Phys. Rev. A {\bf 55} (1997) 4318.
\bibitem{qpm} A. Barone and G. Paterno, {\it Physics and applications of
the Josephson effect} (Wiley, New York, 1982). 
\bibitem{drummond} J.R. Anglin, P. Drummond and A. Smerzi, Phys. Rev. A
{\bf 64} (2001) 063605.
\bibitem{enollskii} V.Z. Enol'skii, M. Salerno, N.A. Kostov and A.C.
Scott, Phys. Scr. {\bf 43} (1991) 229;\\
V.Z. Enol'skii, M. Salerno, A.C. Scott and J.C. Eilbeck, Physica D {\bf
59} (1992) 1. 
\bibitem{us} H.-Q. Zhou, J. Links, R.H. McKenzie and X.-W. Guan, {\it
Exact results for a tunnel-coupled pair of trapped Bose-Einstein 
condensates}, cond-mat/0203009. 
\bibitem{j} B.D. Josephson, Phys. Lett. {\bf 1} (1962) 251;\\
B.D. Josephson, Rev. Mod. Phys. {\bf 46} (1974) 251.
\bibitem{phase} 
S.-X. Yu, Phys. Rev. Lett. {\bf 79} (1997) 780. 
\bibitem{korepin} V.E. Korepin, N.M. Bogoliubov and A.G. Izergin, {\it
Quantum inverse scattering method and correlation functions} (Cambridge
University Press, 1993). 
\bibitem{maillet} N. Kitanine, J.M. Maillet and V. Terras, Nucl. Phys. B
{\bf 554} (1999) 647.
\bibitem{kt} V.B. Kuznetsov and A.V. Tsiganov, J. Phys. A: Math. Gen.
{\bf 22} (1989) L73. 
\bibitem{ik82} A.G. Izergin and V.E. Korepin, Lett. Math. Phys. {\bf 6}
(1982) 283.
\bibitem{slavnov} N.A. Slavnov, Theor. Math. Phys. {\bf 79} (1989) 502.

\end{thebibliography}
\end{document}